\documentclass[a4paper,fleqn,usenatbib,useAMS]{mnras}

\usepackage[T1]{fontenc}
\usepackage{ae,aecompl}
\usepackage{natbib}

\usepackage{graphicx}	
\usepackage{amsmath}	
\usepackage{amssymb}	
\usepackage{multicol}        
\usepackage{bm}		
\usepackage{pdflscape}	
\usepackage{booktabs}
\usepackage{caption}
\usepackage{mathtools}
\usepackage{makecell}

\newcommand{\angstrom}{\textup{\AA}}
\newcommand{\fesc}{\ensuremath{f_{\rm esc}}}
\newcommand{\fout}{\ensuremath{\langle f_{900}/f_{1500}\rangle_{\rm out}}}
\newcommand{\fobs}{\ensuremath{\langle f_{900}/f_{1500}\rangle_{\rm obs}}}

\newcommand{\wlya}{$W_{\lambda}$(Ly$\rm \alpha$)}
\newcommand{\wlis}{$W_{\lambda}$(LIS)}
\newcommand{\whis}{$W_{\lambda}$(HIS)}
\newcommand{\wlyal}{$W_{\lambda}$(Ly$\rm \alpha$)$_{\rm low}$}
\newcommand{\wlyau}{$W_{\lambda}$(Ly$\rm \alpha$)$_{\rm high}$}
\newcommand{\luv}{\ensuremath{L_{\rm UV}}}

\newcommand{\fesca}{\ensuremath{f_{\rm esc,abs}}}

\newcommand{\sfrd}{\ensuremath{\Sigma_{\rm SFR}}}
\newcommand{\sfrdl}{\ensuremath{\Sigma_{\rm SFR,low}}}
\newcommand{\sfrdu}{\ensuremath{\Sigma_{\rm SFR,high}}}
\newcommand{\re}{\ensuremath{r_{\rm e}}}
\newcommand{\vjh}{\ensuremath{V_{606}J_{125}H_{160}}}
\newcommand{\hstj}{\ensuremath{J_{125}}}
\newcommand{\hstv}{\ensuremath{V_{606}}}
\newcommand{\hsth}{\ensuremath{H_{160}}}
\newcommand{\mstar}{\ensuremath{\rm M_{\rm *}}}
\newcommand{\msun}{\ensuremath{M_{\odot}}}

\title[LyC escape and SFR surface density]{Searching for the connection between ionizing-photon escape and the surface density of star formation at $z\sim3$}

\author[Pahl et al.]{Anthony J. Pahl,$^{1}$\thanks{Contact e-mail: \href{mailto:pahl@astro.ucla.edu}{pahl@astro.ucla.edu}}
Alice Shapley,$^{1}$
Charles C. Steidel,$^{2}$
Naveen A. Reddy,$^{3}$\newauthor
and Yuguang Chen$^{2,4}$
	\\
	$^{1}$Department of Physics and Astronomy, University of California, Los Angeles, CA 90095, USA\\
	$^{2}$Cahill Center for Astronomy and Astrophysics, California Institute of Technology, MC249-17, Pasadena, CA 91125, USA\\
	$^{3}$Department of Physics and Astronomy, University of California Riverside, Riverside, CA 92521, USA\\
	$^{4}$Department of Physics and Astronomy, University of California Davis, 1 Shields Avenue, Davis, CA 95616, USA
}

\date{}

\pubyear{}

\begin{document}
		
	\label{firstpage}
	\pagerange{\pageref{firstpage}--\pageref{lastpage}}
	\maketitle
	
\begin{abstract}
	The connection between the escape fraction of ionizing photons ({\fesc}) and star-formation rate surface density ({\sfrd}) is a key input for reionization models, but remains untested at high redshift. 
	We analyse 35 $z\sim3$ galaxies from the Keck Lyman Continuum Survey (KLCS) covered by deep, rest far-UV spectra of the Lyman continuum (LyC) and high-resolution $HST$ {\hstv} imaging, enabling estimates of both {\fesc} and rest-UV sizes. 
	Using Sérsic profile fits to $HST$ images and spectral-energy distribution fits to multi-band photometry, we measure effective sizes and SFRs for the galaxies in our sample, and separate the sample into two bins of {\sfrd}. 
	Based on composite spectra, we estimate $\langle \fesc\rangle$ for both {\sfrd} subsamples, finding no significant difference in $\langle \fesc\rangle$ between the two. 
	To test the representativeness of the KLCS $HST$ sample and robustness of this result, we attempt to recover the well-established correlation between {\fesc} and Ly$\alpha$ equivalent width. 
	This correlation is not significant within the KLCS $HST$ sample, indicating that the sample is insufficient for correlating {\fesc} and galaxy properties such as {\sfrd}.
	We perform stacking simulations using the KLCS parent sample to determine the optimal sample size for robust probes of the {\fesc}-{\sfrd} connection to inform future observing programs.
	For a program with a selection independent of ionizing properties, $\ge90$ objects are required; for one preferentially observing strongly-leaking LyC sources, $\ge58$ objects are required. 
	More generally, measuring the connection between {\fesc} and {\sfrd} requires a larger, representative sample spanning a wide dynamic range in galaxies properties such as {\sfrd}.
\end{abstract}


\defcitealias{Steidel2018}{S18}
\defcitealias{Mostardi2015a}{M15}
\defcitealias{Vanzella2012}{Vanzella et al. 2012}

\section{Introduction} \label{sec:intro}

Reionization is a key phase transition of the Universe in which the neutral hydrogen in the intergalactic medium (IGM) becomes ionized. While the reionization process concludes by $z\sim6$ \citep{Fan2006}, the overall evolution of the hydrogen neutral fraction depends on our understanding of the population of the first star-forming galaxies that are thought to emit the bulk of the ionizing photons into the IGM \citep[e.g,][]{Parsa2018a}. The ionizing emissivity as a function of cosmic time can be simply parameterized as the product of the comoving UV luminosity density ($\rho_{\rm UV}$), ionizing photon efficiency ($\xi_{\rm ion}$), and escape fraction of ionizing photons that go on to ionize the surrounding \ion{H}{I} gas ({\fesc}) \citep{Robertson2015a}. These parameters and their redshift dependence must be characterized beyond $z\sim6$ to construct a complete picture of reionization.

The most difficult of the three parameters to constrain is {\fesc}, for which estimates rely on direct measurement of the Lyman continuum (LyC; $\lambda<912$\AA) spectral region. Due to the large cross section of \ion{H}{I} to LyC photons, even a small neutral fraction $(<1\%)$ can attenuate this signal, thus direct observations become prohibitive at $z\gtrsim4$ \citep{Vanzella2012,Steidel2018}. We must use lower-redshift galaxies to obtain direct measurements of escaping LyC emission at the highest redshift where the measurements are possible, and, simultaneously, connect their properties with those of ionizing sources at $z>6$.
To this end, large-scale LyC surveys have attempted to measure the average {\fesc} of galaxies at $z\sim2-4$ and to understand the mechanisms of LyC escape. These surveys either attempt to detect LyC photometrically \citep{Grazian2017,Fletcher2019,Nakajima2019,Saxena2021} or spectroscopically \citep{Marchi2017,Marchi2018}. Of particular note is the Keck Lyman Continuum Spectroscopic Survey (KLCS), which includes 124 objects selected as Lyman Break Galaxies (LBGs) with $\langle\fesc\rangle=0.09\pm0.01$ measured through deep, rest-UV spectroscopy \citep{Steidel2018}. 
A subset of 39 KLCS galaxies was followed up with \textit{Hubble Space Telescope} ($HST$) multi-band imaging, including all galaxies with apparent individual LyC detections. These observations were designed to eliminate the effects of foreground contamination on the average measurements, a significant concern for detecting LyC leakers at $z\sim3$ \citep{Vanzella2012,Mostardi2015a}. 
In \citet{Pahl2021}, we reported that four KLCS galaxies (two of which had been identified as significant leakers) imaged by $HST$ had multi-band photometry consistent with low-redshift interlopers, and updated the average {\fesc} of the KLCS sample to $0.06\pm0.01$ after removing these four galaxies.

Within these studies, one of the parameters that positively correlates with LyC leakage is the equivalent width of Ly$\alpha$ emission ({\wlya}), in the sense that galaxies with higher {\wlya} have higher {\fesc} on average \citep{Marchi2017,Marchi2018,Steidel2018,Pahl2021}. This correlation suggests that Ly$\alpha$ and LyC leakage in star-forming galaxies are connected because of their similar escape pathways through the interstellar and circumgalactic media (ISM and CGM). Since Ly$\alpha$ emission has been found to be inversely correlated with the neutral gas covering fraction both in the local Universe and at $z\sim3$ \citep{Reddy2016b,Gazagnes2020}, it follows that a ``picket fence" arrangement of \ion{H}{I} would simultaneously modulate LyC emission along with Ly$\alpha$. The galaxy properties that favor this configuration are an active area of study, and further characterization of the dependence of {\fesc} on galaxy properties is necessary to disentangle this physical picture.

Recently, \citet{Naidu2020} proposed a model of reioniziation in which a galaxy's {\fesc} is solely a function of its star-formation rate (SFR) surface density ({\sfrd}). This model implies that the bulk of ionizing photon leakage at $z>6$ can be attributed to relatively luminous and massive galaxies ($M_{\rm UV}<-18$ and log$(M_{*}/M_{\odot})>8$), in stark contrast to models that favor the dominant role of dwarf galaxies at $M_{\rm UV}>-15$ \citep[e.g.,][]{Finkelstein2019}.
The parameterization of {\fesc} as a function of {\sfrd} is motivated by the above-average {\sfrd} found in many individual LyC detections at $z\sim2-4$ \citep{DeBarros2016,Vanzella2016,Shapley2016a,Bian2017,Vanzella2017a}, as well as in the $HST$/COS sample of LyC detections at $z\sim0.3$ \citep{Borthakur2014,Izotov2016a,Izotov2018,Izotov2021}. Additionally, the survey of \citet{Marchi2018} found that stacks of galaxies with small UV spatial size had a significantly higher average LyC flux than those with larger size, implying that compact galaxies have conditions favorable for LyC escape. The spectroscopic survey of \citet{Reddy2021} found Ly$\alpha$ escape can be aided by a high {\sfrd}, especially in low stellar mass galaxies. Finally, hydrodynamic simulations including radiative transfer, such as those of \citet{Ma2016}, predict that feedback from massive stars carves channels in the ISM that allow for efficient leakage of LyC. This effect is predicted to be most pronounced in galaxies containing a high density of star formation. The {\fesc}-{\sfrd} paradigm has important implications for testing the model of reionization presented by \citet{Naidu2020}, and for informing the evolution of the \ion{H}{I} fraction, but has not been confirmed observationally in a large-scale survey with selection criteria independent of {\sfrd}.

In this work, we attempt the first LBG-selected study of {\fesc} as a function of {\sfrd} using the subset of KLCS galaxies covered by $HST$ imaging. Thanks to the high-resolution $HST$ imaging offered by \citet{Pahl2021}, well-constrained size measurements are possible for the first time for 35 of the galaxies in KLCS, including the 13 with significant individual LyC detections. Combined with the ability to estimate SFRs from the wealth of multi-band photometry available for KLCS, this subsample is ideal for the first observational test of of the connection between ionizing photon escape and {\sfrd}. Using the deep, rest-frame UV spectra of KLCS, which include coverage of LyC at rest-frame $880-910\textrm{\AA}$ at $z\sim3$, these spectra can be stacked as a function of {\sfrd} to explore characteristics of the LyC spectral region. While stacking reduces the uncertainty due to line-of-sight variations in IGM transmission, it remains to investigate whether this KLCS $HST$ subsample is adequate to constrain any dependence of {\fesc} on {\sfrd}.

The structure of the paper is as follows: in Section \ref{sec:sample}, we describe the observations and data available for the KLCS $HST$ sample. In Section \ref{sec:methods}, we describe the SFR and size measurements, which we present in Section \ref{sec:res} alongside descriptions and measurements of the composite spectra constructed in two bins of {\sfrd}. In Section \ref{sec:disc}, we examine the robustness of the KLCS $HST$ sample for quantifying the correlation between {\fesc} and {\sfrd}, and perform simulations to determine the minimum sample composition for analyzing the relationship between the escape of ionizing radiation and galaxy properties. We summarize our conclusions in Section \ref{sec:summary}.

Throughout this paper, we adopt a standard $\Lambda$CDM cosmology with $\Omega_m$ = 0.3, $\Omega_{\Lambda}$ = 0.7 and $H_0$ = 70 $\textrm{km\,s}^{-1}\textrm{Mpc}^{-1}$. The {\fesc} values reported in this paper are absolute escape fractions, equivalent to {\fesca} in \citet{Steidel2018}, and defined as the fraction of all H-ionizing photons produced within a  galaxy that escapes into the IGM. We also employ the AB magnitude system \citep{Oke1983}.

\section{Sample} \label{sec:sample}

The KLCS \citep{Steidel2018} includes an analysis of the hydrogen-ionizing spectra of 137 galaxies selected as LBGs \citep[for details of selection see][]{Steidel2003,Adelberger2004,Reddy2012}. To this end, deep optical spectra were obtained using the Low-Resolution Imaging Spectrometer \citep[LRIS;][]{Oke1995,Steidel2004} on the Keck I telescope, covering the LyC spectral region, as well as the Ly$\alpha$ feature and far-UV metal absorption lines. These data were taken with total exposure times per mask of $\sim10\:$hrs. Thirteen galaxies were removed from the sample due to clear evidence of blending, the presence of instrumental defects, or with evidence of multiple spectroscopic redshifts, for a final analysis sample of 124 objects. \citet{Steidel2018} characterized the escaping ionizing signal from each galaxy using the average flux density in the wavelength region $880\textrm{\AA}-910\textrm{\AA}$, or $f_{900}$. Fifteen galaxies with significant LyC escape (defined as $f_{900}>3\sigma_{900}$, where $\sigma_{900}$ is the error on the flux density in the LyC wavelength region) were defined as the LyC detection sample. The remaining 109 galaxies were characterized as individual LyC non-detections. Despite the best efforts to clean the sample of interloper contamination, high-resolution imaging is necessary for identifying all low-redshift interlopers along the line-of-sight that frequently contaminate apparent LyC flux for $z\sim3$ galaxies \citep{Vanzella2012,Mostardi2015a}.

In \citet{Pahl2021}, the 15 LyC detection candidates from KLCS were followed up as part of a Cycle 25 $HST$ Program ID 15287 (PI: Shapley) in order to explore the effects of foreground contamination in the KLCS sample. While these LyC detection candidates were specifically targeted as the most likely sources of contamination due to their significant individual $f_{900}$ measurements, 24 LyC individual non-detections were also observed as they fell within the field of view of the $HST$ pointings. These 39 objects were observed with ACS/F606W ($V_{606}$), WFC3/F125W ($J_{125}$), and WFC3/F160W ($H_{160}$) between 2017 and 2019 in five KLCS survey fields (Q0933, Q1422, DSF2237b, Q1549, and Westphal). Additional $HST$ data were included from \citet{Mostardi2015a} and \citet{Shapley2016a} for one object, Q1549-C25. The seven ACS pointings and 11 WFC3 pointings were observed for three orbits per filter. We used {\vjh} photometry to characterize the contamination likelihood of individual subcomponents within the LyC detection sample, following the methodology of \citet{Mostardi2015a}. To this end, two LyC detections were determined to be likely contaminated and were subsequently removed from the sample. Similar contamination analysis was performed on the objects with individual LyC non-detections for completeness. Two galaxies with individual non-detections had subcomponents with {\vjh} photometry consistent with low-redshift interlopers, thus these objects were also removed, for a remaining KLCS $HST$ sample of 35.

We use this ``KLCS $HST$" sample of 35 objects that were determined to be free of contamination for the analysis presented here. While all KLCS objects have multi-band photometry that can be used to estimate the SFR through spectral-energy distribution (SED) fitting, only those targeted by $HST$ have the higher resolution imaging necessary to measure accurate rest-UV sizes. In this KLCS $HST$ sample, there are 13 galaxies with individual LyC detections, and 22 galaxies with individual LyC non-detections, as defined by \citet{Steidel2018}. Thus, the KLCS $HS$T subsample sample is biased in favor of LyC detections compared to the full sample, which contains 13 individual LyC detections out of a total of 120 galaxies. We discuss the effects of this selection bias in Section \ref{sec:res}. 

\section{Methodology} \label{sec:methods}

In this section, we present the steps necessary to measure {\sfrd} for the KLCS $HST$ sample. This measurement requires sizes, estimated by fitting the two-dimensional light profiles in \textit{HST} imaging, and SFRs, calculated from SED fitting.

\subsection{Sizes} \label{sec:re}

We estimated the sizes of objects in our sample from model fits to the light distributions in the {\hstv} images, which have the highest resolution within our $HST$ dataset. For this modeling, we used the software \textsc{galfit} \citep{Peng2002, Peng2010}, which uses parametric fits of one or more Sérsic profiles and a uniform sky background to the two-dimensional postage stamp of a galaxy. At the redshifts of our sample galaxies, single Sérsic profiles have been demonstrated to reasonably constrain the effective sizes of distinct morphological components in $HST$ imaging \citep{VanDerWel2012, Law2011, Shibuya2015, Gillman2020}. We follow \citet{VanDerWel2012}, who presented robust measurements using Sérsic profiles  of basic morphological properties such as half-light radii for galaxies in the CANDELS survey down to $H_{160}\sim24.5$.

The free parameters of the fits are central position (\textit{x}, \textit{y}), magnitude, Sérsic index (\textit{n}), position angle, axis ratio ($b/a$), and half-light radius along the major axis (\textit{r}) for each profile, as well as a uniform sky background level. One of the inputs to the \textsc{galfit} modeling is a bad-pixel mask for each object based on the segmentation map produced by \textit{SExtractor} \citep{Bertin1996}. The \textit{SExtractor} detection threshold  was set at 2.5 times the background RMS, and the background was estimated using the ``local" setting, which estimates the sky contribution using a rectangular annulus. All components unrelated to the central object were masked, ensuring that the flux from surrounding sources does not affect the fit. Deconvolution was performed using point-spread functions (PSFs) described in \citet{Pahl2021}. These PSFs were constructed for each field by averaging $10-15$ unsaturated stars in the {\hstv} mosaic using the \textit{Astropy} routine \textit{Photutils} \citep{Bradley2020}.

\begin{figure*} 
	\centering
	\includegraphics[width=0.94\textwidth]{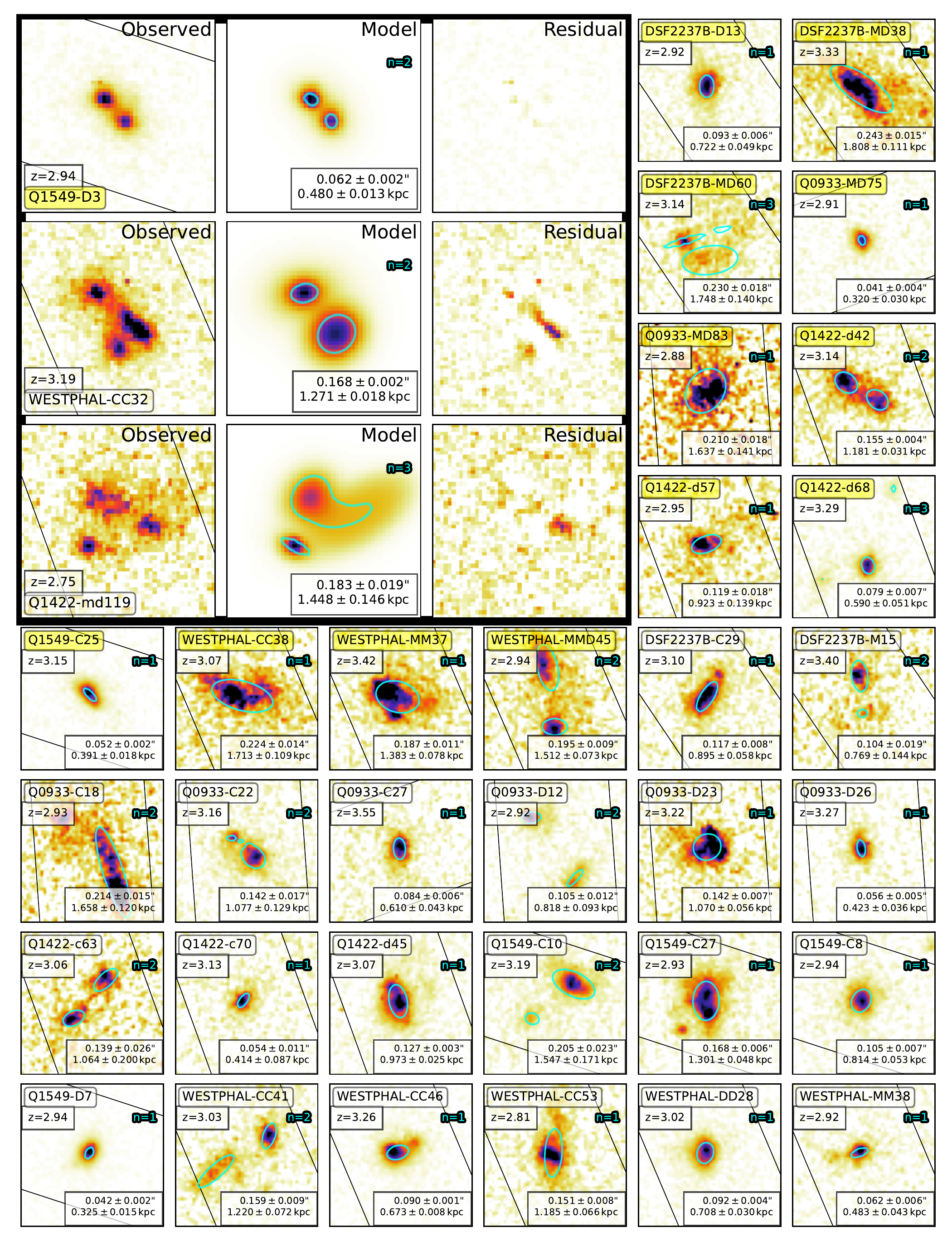}
	\caption{$V_{606}$ postage stamps of the 35 galaxies in KLCS $HST$. The panels in the upper left show the HST {\hstv} image, the \textsc{galfit} model, and the residual, from left to right, for three objects to demonstrate the fitting process. The cyan curve shows the effective area of each object and the black lines show the LRIS slit (1.2$\arcsec$ wide) for each object. Displayed in the bottom right of each panel is the circularized {\re} in both angular size ($\arcsec$) and physical (kpc) size. Displayed in the middle right of each panel in cyan are the number of Sérsic profiles used in the model fit. Objects with individual LyC detections have object names highlighted in yellow. All panels are $1.5\arcsec$ per side.
	}
	\label{fig:galfit}
\end{figure*}

We initialized the positions of the Sérsic profiles for each object using the central coordinates of up to three of the brightest components deblended with \textit{SExtractor}. We set an initial guess of \textit{n}=4 and \textit{r}=5 pixels. The initial guesses for the magnitude of each source and the sky background level were drawn from the \textit{SExtractor} catalogs. After fitting, the residuals were inspected by eye to ensure that the background levels were properly subtracted and the models accurately described the radial extent of the light profile. In the top three panels of Figure \ref{fig:galfit}, the fitting process is detailed for the galaxy Q1549-D3, with the $HST$ $V_{606}$ image, model fit, and residual displayed from left to right.

To estimate an effective size for objects with single-component fits, we calculated the circularized effective radius ({\re}) by combining the best-fit \textit{r} and $b/a$ parameters from \textsc{galfit} according to $\re=r\sqrt{b/a}$. We defined the effective area of the galaxy as $\pi\re^2$, which corresponds to the projected area containing half of the total light of the galaxy.

For objects with multi-component fits, we estimated the effective area using the best-fit model produced by \textsc{galfit}. 
By using the best-fit model without the PSF convolution, we effectively recreated the deconvolved image for each object.
We integrated this deconvolved model to find a total flux $F$, then sorted the pixels in brightness. A number $n_{\rm pix}$ of the brightest pixels were chosen, starting at the brightest pixel and continuing towards fainter pixels, such that the integrated flux contained within $n_{\rm pix}$ was $F/2$. The effective area of an object was then found by applying the $HST\:V_{606}$ pixel scale of $0.03\arcsec/\rm pix$, which we subsequently converted to a circularized radius using $n_{\rm pix}\times(0.03(\arcsec/\textrm{pix}))^2 = \pi\re^2$.

We pursued an empirical approach for estimating the errors on the effective sizes of the galaxies in our sample. We randomly perturbed each pixel of the science image by a value drawn from a Gaussian distribution with standard deviation corresponding to the pixel value in the error image generated from the reduction process of the $HST$ {\hstv} imaging \citep[See Section 3.2;][]{Pahl2021}. To ensure our error bars encapsulated any systematic error induced by different initial guesses, we also randomly perturbed the initial guesses of all model parameters. Starting with the best-fit parameters from the unperturbed fit, we perturbed the \textit{x} and \textit{y} positions by a Gaussian distribution with width 2 pixels, and multiplied \textit{r} by a factor drawn from a Gaussian distribution with a mean of 1.0 and a standard deviation of 0.3. We randomized \textit{n} uniformly between 0.5 and 4.0, and $a/b$ between 0.0 and 1.0. We then fit the perturbed science image with these randomized initial guesses following the methods described above and subsequently measured {\re}. After 100 realizations, the standard deviation of the {\re} distribution was taken as the error on the object's {\re} measurement. These empirically-determined errors are comparable to those calculated internally by \textsc{galfit} for the objects with single-Sérsic fits.

We present the postage stamps and sizes of the 35 KLCS HST objects in Figure \ref{fig:galfit}. Displayed in cyan are curves containing the effective areas: for single-component morphologies, these are defined by the Sérsic model parameters \textit{r} and $a/b$, and, for multi-component morphologies, the cyan curves enclose the pixels corresponding to the half-light area as described above. In the bottom-right corner of each panel, we present the circularized {\re} for each object in both angular ($\arcsec$) and physical (kpc) size. Displayed in black is the LRIS slit for each object: all components that contribute to the {\re} measurement are contained within the slit for KLCS $HST$ targets.

\subsection{Star-formation rates} \label{sec:sfr}

We fit multi-band photometry available for our objects with stellar population synthesis models in order to derive galaxy properties such as stellar mass ({\mstar}), $E(B-V)$, stellar age, and SFR. While the SFR measurement is the focus of this work, we will explore other derived galaxy properties of KLCS in future work. 

We measured {\vjh} photometry for the KLCS $HST$ objects based on $HST$ images reduced and described in \citet{Pahl2021}. We followed a similar measurement procedure to that of \citet{Pahl2021}, however, for consistency with the ground-based measurements, we set the detection threshold used with \textit{SExtractor} to 2.5 times the background RMS, and removed any deblending to get singular measurements for each object.

For additional input photometry that constitute the KLCS $HST$ SEDs, we used a multitude of data available for the objects in the survey fields \citep[e.g.,][]{Reddy2012}.
At minimum, all five fields were covered by ground-based $U_n G R_s$ optical, $K_s$ near-IR, and $Spitzer$/IRAC Channel 1 and 2 infrared imaging. Some fields had additional photometric information, as summarized in Table \ref{tab:obs}. For objects with a clear $K_s$ excess, we performed corrections to the broadband flux based on measured fluxes of the strong emission lines of [OIII] and H$\alpha$ \citep{Steidel2014}.

\begin{table*}
	\centering
	\caption{Photometric bands used in SED modeling.}
	\begin{tabular}{cc}
		\toprule
		Fields   & Photometric bands                                         \\ \midrule
		\midrule
		DSF2237b & $U_n^a$, $G^a$, $R_s^a$, $I^a$, $J^b$, $K_s^b$, IRAC1, IRAC2, {\hstv}, {\hstj}, {\hsth}  \\ \midrule
		Q0933    & $U_n^a$, $G^a$, $R_s^a$, $I^a$, $J^b$, $K_s^b$, IRAC1, IRAC2, {\hstv}, {\hstj}, {\hsth}         \\ \midrule
		Q1422    & $U_n^c$, $G^c$, $R_s^c$, $K_s^b$, IRAC1, IRAC2, {\hstv}, {\hstj}, {\hsth}            \\ \midrule
		Q1549    & $U_n^d$, $G^d$, $R_s^d$, $J^f$, $H^f$, $K_s^{b,f}$, IRAC1, IRAC2, IRAC3, {\hstv}, {\hstj}, {\hsth}        \\ \midrule
		Westphal & u*$^g$, g'$^g$, r'$^g$, i'$^g$, z'$^g$, $J^h$, $H^h$, $K_s^h$, IRAC1, IRAC2, {\hstv}, {\hstj}, {\hsth}        \\ \bottomrule
	\end{tabular}
	\begin{flushleft}
		$^a$ {Observed with the COSMIC prime focus imager on the Palomar 5.08 m telescope \citep[see][]{Steidel2003}.}
		$^b$ {Observed with the Multi-Object Spectrograph for Infra-Red Exploration (MOSFIRE) on the Keck I telescope.}
		$^c$ {Observed with the Prime Focus Imager on the William Herschel 4.2m  telescope (WHT) \citep[see][]{Steidel2003}.}
		$^d$ {Observed with Keck/LRIS.}
		$^f$ {Observed with FourStar at the Magellan Baade 6.5m telescope.}
		$^g$ {From the Canada-France-Hawaii Telescope (CFHT) Legacy Survey.}
		$^h$ {Observed with CFHT/WIRCam as part of the WIRCam Deep Survey \citep{Bielby2012}.}
	\end{flushleft}
\label{tab:obs}
\end{table*}

We measured the SFRs for each object using SED fitting based on the methods described in \citet{Reddy2018}. Specifically, we used \citet{Bruzual2003} stellar population synthesis models and assumed a \citet{Salpeter1955} initial mass function (IMF). We note that, using the stellar population synthesis models of BPASS \citep[BPASS v2.1,][]{Eldridge2017} and assuming the same IMF, we find excellent agreement for the inferred SFRs. A constant star-formation history with stellar ages no younger than $50\:\textrm{Myr}$ was assumed, such that stellar ages would not extend to younger ages than the typical dynamical timescales of star-forming galaxies at $z\sim3$ \citep{Reddy2012}. We fit the SEDs using two sets of assumptions for metallicity and dust attenuation curve: 0.28 times solar metallicity and an SMC extinction curve \citep[i.e., 0.28Z$_{\odot}$+SMC,][]{Gordon2003}, and 1.4 times solar metallicity and a \citet{Calzetti1999} extinction curve (i.e., 1.4Z$_{\odot}$+Calzetti). For two galaxies with log($\mstar/\msun$)>10.65 measured using the 0.28Z$_{\odot}$+SMC models, we adopted the 1.4Z$_{\odot}$+Calzetti model for the final fit, based on systematically better fits to the observed SEDs in \citet{Du2018} and for consistency with the mass-metallicity relation \citep{Sanders2015, Steidel2014, Onodera2016}. With the exception of the fits to these two galaxies, we adopted 0.28Z$_{\odot}$+SMC models.

To ensure robust SED fits, we individually examined the input photometry, best-fit model, and resulting unreduced $\chi^2$ for each object. Due to the higher S/N of $HST$, we removed MOSFIRE or WIRCAM $J$, $H$, and $K_s$ data that were inconsistent with $HST\:\hstj$, as long as the IRAC photometry constrained the SED redward of the Balmer break. We also removed IRAC photometry that appeared to be blended with other nearby sources. Additionally, we removed any ground-based photometry if the ground-based extraction may have included contamination from nearby objects identified with the higher-resolution $HST$ imaging. After eliminating clearly problematic photometry, we achieved SED fits with unreduced $\chi^2<30$ for all 35 objects.

\section{Results} \label{sec:res}

In this section, we present the {\sfrd} measurements for the KLCS $HST$ sample and the ionizing and non-ionizing spectral properties of the sample as a function of {\sfrd}.

\begin{figure*} 
	\centering
	\includegraphics[width=\textwidth]{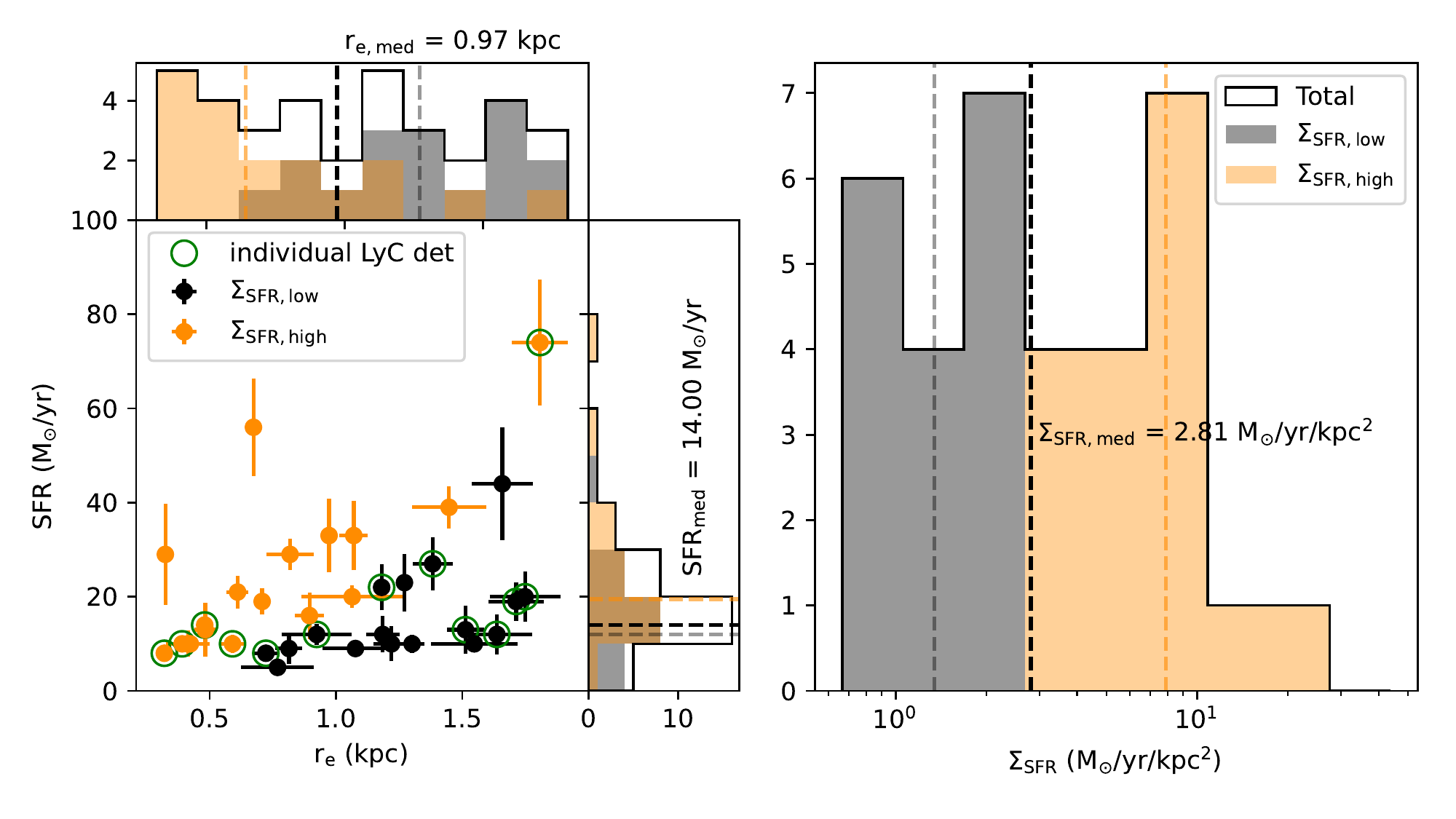}
	\caption{{\re}, SFR, and {\sfrd} measurements for the KLCS $HST$ sample. The 35 objects are split into two bins of {\sfrd}: {\sfrdl} contains the 17 objects with the lowest {\sfrd}, and {\sfrdu} contains the 18 objects with the highest {\sfrd}. Objects in the {\sfrdl} bin are displayed as black points in the scatter plot  and black bars in the histograms, while objects in the {\sfrdu} bin are displayed as orange points and orange bars. The thirteen objects with individual LyC detections are highlighted with a green circle.
		\textbf{Left:} The SFR-{\re} distribution of the sample. The dark, black line in the vertical and horizontal histograms show the total {\re} and SFR distributions. The sample medians are shown by dashed lines, with the {\sfrdl} medians in grey, the {\sfrdu} medians in orange, and the total sample median in black.
		\textbf{Right:} The {\sfrd} distribution of the sample, calculated according to Equation \ref{eqn:sfrd}. The color convention follows the histograms in the left panel.
	}
	\label{fig:sfrd}
\end{figure*}

We present the {\re} and SFR measurements for our sample in the left panel of Figure \ref{fig:sfrd}. The median {\re} is $0.97\:\textrm{kpc}$ (for a median $M_{\rm UV}=-20.7$), comparable to $1.054^{+1.154}_{-0.436}\:\textrm{kpc}$ found for $z\sim3-4$ star-forming galaxies at $M_{\rm UV}=-21.0$ \citep{Shibuya2015}.
The median SFR of the sample is 14.0 $\msun\,\textrm{yr}^{-1}$. For each object, we calculate the {\sfrd} by assuming half of the star formation occurs within the half-light area, such that
\begin{equation} \label{eqn:sfrd}
 \sfrd=\textrm{SFR}/(2\times\pi\re^2),
\end{equation}
where SFR is inferred from the best-fit SED model described in Section \ref{sec:sfr}, and {\re} is the circularized half-light radius measured for each object in Section \ref{sec:re}.
We present the resulting {\sfrd} distribution in the right panel of Figure \ref{fig:sfrd}. The median {\sfrd} is $2.81\:\msun\,\textrm{yr}^{-1}\,\textrm{kpc}^{-2}$, which is within the 16th and 84th percentile of the {\sfrd} relation from \citet{Shibuya2015} with comparable redshift and $M_{\rm UV}$.

In order to examine the ionizing and non-ionizing spectral properties of the KLCS $HST$ sample as a function of {\sfrd}, we split the sample into two bins: {\sfrdl}, which contains the 17 galaxies with the lowest {\sfrd}, and {\sfrdu}, which contains the 18 galaxies with the highest {\sfrd}. The two bins are differentiated by black and orange color for {\sfrdl} and {\sfrdu}, respectively, in Figure \ref{fig:sfrd}. The median properties of the samples are summarized in Table \ref{tab:prop}.

\begin{table*}
\centering
\caption{Median properties of the two {\sfrd} subsamples and resulting composite spectra.}
\begin{tabular}{lllllllll}
	\toprule
	{} & \thead{$r_{\rm e,med}$\\(kpc)} & \thead{SFR$_{\rm med}$\\$(\textrm{M}_{\odot}\textrm{yr}^{-1})$} & \thead{$\Sigma_{\rm SFR,med}$\\$(\textrm{M}_{\odot}\textrm{yr}^{-1}\textrm{kpc}^{-2}$)} & \thead{$W_{\lambda}$(Ly$\rm \alpha$)\\(\AA)} & \thead{$W_{\lambda}$(LIS)\\(\AA)} & \thead{$W_{\lambda}$(HIS)\\(\AA)} & $\langle f_{900}/f_{1500}\rangle_{\rm out}$ & $f_{\rm esc,abs}$ \\
	\midrule
	$\sfrdl$ &                  $1.27\pm0.11$ &                                       $12.0\pm2.4$ &                                      $1.35\pm0.40$ &                                 $10.7\pm4.9$ &                     $1.78\pm0.21$ &                     $2.81\pm0.32$ &                             $0.104\pm0.013$ &   $0.120\pm0.009$ \\
	$\sfrdu$ &                  $0.64\pm0.12$ &                                       $19.5\pm4.6$ &                                      $7.89\pm1.53$ &                                $19.1\pm10.8$ &                     $1.66\pm0.20$ &                     $3.20\pm0.32$ &                             $0.076\pm0.013$ &   $0.122\pm0.010$ \\
	\bottomrule
\end{tabular}
\label{tab:prop}
\end{table*}

We then created composite spectra as described in \citet{Steidel2018} and \citet{Pahl2021} in order to explore spectral differences between the {\sfrdl} and {\sfrdu} samples and constrain the dependence of LyC ionizing photon escape on {\sfrd}. The rest-frame UV, LRIS spectra of \citet{Steidel2018} covering the LyC region are normalized to the average flux density within the spectral region $1475 \leq \lambda_{0}/\angstrom \leq 1525$ ($f_{1500}$). We use a spline interpolation to place the spectra on a common wavelength grid, then use a mean combination at each wavelength increment to create the final composite after applying $3\sigma$ outlier rejection. 

Building spectral composites smooths out variations in intergalactic medium (IGM) sight lines. However, the aggregate effect of the IGM and circumgalactic medium (CGM) strongly attenuates the LyC region at $z\sim3$. To account for this attenuation, we correct the composites using IGM+CGM transmission functions calculated in a manner consistent with that in \citet{Steidel2018}. Notably, for the KLCS $HST$ objects, we choose a transmission function drawn from the third quartile of IGM transparency, as opposed to the average transmission used for the composites of \citet{Steidel2018}.  This decision is based on the oversampling of objects with individual LyC detections in the KLCS $HST$ sample (13 detections out of 35 objects) compared to the full KLCS sample (13 detections out of 120 objects, see Section \ref{sec:sample}).
Preferentially targeting individually-detected LyC candidates has the approximate effect of choosing IGM sight-lines that are more transparent to LyC photons, on average, considering that part of the variation in observed ionizing signal at $z\sim3$ is due to differences in IGM opacity \citep{Steidel2018}. Assuming the LyC individual detections are drawn from the top 11\% of IGM transmission\footnote{The 13 LyC individual detections had the highest $f_{900}$ out of the parent sample of 120.} and the 22 LyC non-detections have average sight-lines, we consider the third-quartile transmission function as a reasonable approximation of the IGM transmission for the KLCS $HST$ sample bins. Below, we discuss the results based on a different assumption for IGM transmission, demonstrating that our key conclusions do not change.

The {\sfrdl} and {\sfrdu} composite spectra, corrected for IGM+CGM attenuation, are shown in Figure \ref{fig:spec}. The LyC region ($880 \leq \lambda_{0}/\angstrom \leq 910$) is visually demarcated by dotted-red lines.
We calculated the average ionizing flux density of each composite within this region and defined it as $\langle f_{900}\rangle$. The ratio of ionizing to non-ionizing flux density, with $f_{900}$ corrected for effects of the IGM and CGM, is $\fout=0.104\pm0.013$ for the {\sfrdl} composite and $\fout=0.076\pm0.013$ for the {\sfrdu} composite. Both of these {\fout} measurements are significantly elevated from the full, uncontaminated KLCS sample measurement of $\fout=0.015\pm0.02$ \citep{Pahl2021}. Again, this difference is expected due to the larger proportion of individually-detected LyC galaxies in KLCS $HST$. We summarize the sample composition of each {\sfrd} bin in Table \ref{tab:gals} and the properties of each KLCS $HST$ galaxy. In {\sfrdl}, 8/17 galaxies are LyC individual detections and in {\sfrdu}, 5/18 galaxies are individually detected.

\begin{table*}
	\centering
	\caption{Properties of the KLCS $HST$ sample, split into two bins of {\sfrd}.}
	\begin{tabular}{lllllclr}
		\toprule
		{} & $z_{\rm sys}$ & \thead{$r_{\rm e}$\\(kpc)} & \thead{SFR\\$(\textrm{M}_{\odot}\textrm{yr}^{-1})$} & \thead{$\Sigma_{\rm SFR}$\\$(\textrm{M}_{\odot}\textrm{yr}^{-1}\textrm{kpc}^{-2}$)} & \thead{$W_{\lambda}$(Ly$\rm \alpha$)\\(\AA)} & $L_{\rm UV}/L_{\rm UV}^{*}$ & $\langle f_{900}/f_{1500}\rangle_{\rm obs}$ \\
		\midrule
		\multicolumn{8}{c}{$\sfrdl$} \\ 
		\midrule
		Q1549-C10      &      $3.1894$ &              $1.55\pm0.17$ &                                       $10.0\pm1.5$ &                                      $0.66\pm0.18$ &                                       $23.2$ &                      $0.75$ &                             $0.022\pm0.016$ \\
		\textbf{Q0933-MD83}     &      $2.8800$ &              $1.64\pm0.14$ &                                       $12.0\pm4.3$ &                                      $0.71\pm0.28$ &                                        $3.8$ &                      $0.60$ &                             $0.150\pm0.041$ \\
		\textbf{Westphal-MMD45} &      $2.9357$ &              $1.51\pm0.07$ &                                       $13.0\pm5.1$ &                                      $0.90\pm0.37$ &                                      $-19.1$ &                      $1.42$ &                             $0.086\pm0.023$ \\
		Q1549-C27      &      $2.9256$ &              $1.30\pm0.05$ &                                       $10.0\pm1.9$ &                                      $0.94\pm0.19$ &                                       $14.5$ &                      $0.88$ &                            $-0.014\pm0.019$ \\
		\textbf{Westphal-CC38}  &      $3.0729$ &              $1.71\pm0.11$ &                                       $19.0\pm4.1$ &                                      $1.03\pm0.26$ &                                        $8.6$ &                      $1.00$ &                             $0.056\pm0.013$ \\
		\textbf{DSF2237b-MD60}  &      $3.1413$ &              $1.75\pm0.14$ &                                       $20.0\pm5.3$ &                                      $1.04\pm0.32$ &                                        $7.2$ &                      $0.67$ &                             $0.090\pm0.018$ \\
		Westphal-CC41  &      $3.0268$ &              $1.22\pm0.07$ &                                       $10.0\pm3.7$ &                                      $1.07\pm0.42$ &                                        $9.5$ &                      $0.62$ &                            $-0.031\pm0.022$ \\
		Q0933-C22      &      $3.1639$ &              $1.08\pm0.13$ &                                        $9.0\pm0.7$ &                                      $1.23\pm0.31$ &                                       $57.6$ &                      $0.73$ &                            $-0.008\pm0.020$ \\
		DSF2237b-M15   &      $3.4034$ &              $0.77\pm0.14$ &                                        $5.0\pm0.9$ &                                      $1.35\pm0.55$ &                                        $6.7$ &                      $0.65$ &                            $-0.010\pm0.031$ \\
		Westphal-CC53  &      $2.8070$ &              $1.19\pm0.07$ &                                       $12.0\pm3.9$ &                                      $1.36\pm0.46$ &                                      $-12.3$ &                      $0.49$ &                             $0.013\pm0.029$ \\
		Q1549-C8       &      $2.9373$ &              $0.81\pm0.05$ &                                        $9.0\pm3.3$ &                                      $2.16\pm0.85$ &                                      $-10.4$ &                      $0.54$ &                             $0.017\pm0.030$ \\
		\textbf{Q1422-d57}      &      $2.9461$ &              $0.92\pm0.14$ &                                       $12.0\pm2.2$ &                                      $2.24\pm0.79$ &                                       $52.8$ &                      $0.30$ &                             $0.294\pm0.103$ \\
		\textbf{Westphal-MM37}  &      $3.4215$ &              $1.38\pm0.08$ &                                       $27.0\pm5.6$ &                                      $2.25\pm0.53$ &                                        $7.4$ &                      $1.23$ &                             $0.046\pm0.009$ \\
		Westphal-CC32  &      $3.1924$ &              $1.27\pm0.02$ &                                       $23.0\pm6.2$ &                                      $2.27\pm0.61$ &                                       $15.6$ &                      $1.40$ &                             $0.009\pm0.009$ \\
		\textbf{DSF2237b-D13}   &      $2.9216$ &              $0.72\pm0.05$ &                                        $8.0\pm1.6$ &                                      $2.44\pm0.58$ &                                       $-3.8$ &                      $0.58$ &                             $0.077\pm0.019$ \\
		\textbf{Q1422-d42}      &      $3.1369$ &              $1.18\pm0.03$ &                                       $22.0\pm4.9$ &                                      $2.51\pm0.57$ &                                      $-10.5$ &                      $0.47$ &                             $0.142\pm0.038$ \\
		Q0933-C18      &      $2.9261$ &              $1.66\pm0.12$ &                                      $44.0\pm12.0$ &                                      $2.55\pm0.78$ &                                      $-13.0$ &                      $0.76$ &                             $0.015\pm0.030$ \\
		\midrule
		\multicolumn{8}{c}{$\sfrdu$} \\ 
		\midrule
		Q1422-c63      &      $3.0591$ &              $1.06\pm0.20$ &                                       $20.0\pm2.4$ &                                      $2.81\pm1.11$ &                                       $-0.1$ &                      $0.28$ &                            $-0.005\pm0.069$ \\
		Q1422-md119    &      $2.7506$ &              $1.45\pm0.15$ &                                       $39.0\pm4.5$ &                                      $2.96\pm0.69$ &                                       $-5.0$ &                      $0.51$ &                             $0.084\pm0.084$ \\
		DSF2237b-C29   &      $3.0999$ &              $0.90\pm0.06$ &                                       $16.0\pm5.0$ &                                      $3.18\pm1.07$ &                                       $27.8$ &                      $0.64$ &                            $-0.036\pm0.018$ \\
		\textbf{DSF2237b-MD38}  &      $3.3278$ &              $1.81\pm0.11$ &                                      $74.0\pm13.3$ &                                      $3.60\pm0.78$ &                                      $-13.8$ &                      $1.47$ &                             $0.075\pm0.009$ \\
		\textbf{Q1422-d68}      &      $3.2865$ &              $0.59\pm0.05$ &                                       $10.0\pm2.0$ &                                      $4.57\pm1.19$ &                                      $153.4$ &                      $0.88$ &                             $0.200\pm0.028$ \\
		Q0933-D23      &      $3.2241$ &              $1.07\pm0.06$ &                                       $33.0\pm7.3$ &                                      $4.58\pm1.13$ &                                      $-14.9$ &                      $1.29$ &                             $0.003\pm0.016$ \\
		Q1422-d45      &      $3.0717$ &              $0.97\pm0.02$ &                                       $33.0\pm7.9$ &                                      $5.55\pm1.35$ &                                        $0.3$ &                      $1.38$ &                                     $0.000$ \\
		Westphal-DD28  &      $3.0206$ &              $0.71\pm0.03$ &                                       $19.0\pm2.8$ &                                      $6.04\pm1.02$ &                                      $-14.6$ &                      $1.62$ &                            $-0.036\pm0.017$ \\
		Q0933-D12      &      $2.9242$ &              $0.82\pm0.09$ &                                       $29.0\pm3.3$ &                                      $6.90\pm1.76$ &                                        $3.5$ &                      $1.13$ &                             $0.028\pm0.025$ \\
		Westphal-MM38  &      $2.9247$ &              $0.48\pm0.04$ &                                       $13.0\pm5.7$ &                                      $8.88\pm4.19$ &                                       $-7.0$ &                      $0.64$ &                                    $-0.000$ \\
		Q0933-D26      &      $3.2662$ &              $0.42\pm0.04$ &                                       $10.0\pm1.2$ &                                      $8.88\pm1.87$ &                                       $16.0$ &                      $0.87$ &                             $0.052\pm0.027$ \\
		Q0933-C27      &      $3.5463$ &              $0.61\pm0.04$ &                                       $21.0\pm3.5$ &                                      $8.97\pm1.94$ &                                       $46.6$ &                      $0.59$ &                            $-0.042\pm0.021$ \\
		Q1422-c70      &      $3.1286$ &              $0.41\pm0.09$ &                                       $10.0\pm2.8$ &                                      $9.28\pm4.66$ &                                        $6.3$ &                      $0.42$ &                             $0.020\pm0.056$ \\
		\textbf{Q1549-D3}       &      $2.9373$ &              $0.48\pm0.01$ &                                       $14.0\pm1.7$ &                                      $9.66\pm1.31$ &                                        $8.7$ &                      $1.16$ &                             $0.056\pm0.009$ \\
		\textbf{Q1549-C25}      &      $3.1526$ &              $0.39\pm0.02$ &                                       $10.0\pm1.3$ &                                     $10.41\pm1.67$ &                                       $16.6$ &                      $0.74$ &                             $0.083\pm0.019$ \\
		\textbf{Q0933-MD75}     &      $2.9131$ &              $0.32\pm0.03$ &                                        $8.0\pm0.9$ &                                     $12.40\pm2.74$ &                                       $76.9$ &                      $0.89$ &                             $0.095\pm0.023$ \\
		Westphal-CC46  &      $3.2608$ &              $0.67\pm0.01$ &                                      $56.0\pm10.3$ &                                     $19.66\pm3.66$ &                                        $1.2$ &                      $2.33$ &                             $0.004\pm0.006$ \\
		Q1549-D7       &      $2.9362$ &              $0.32\pm0.01$ &                                      $29.0\pm10.7$ &                                    $43.82\pm16.63$ &                                       $11.5$ &                      $1.07$ &                             $0.001\pm0.018$ \\
		\bottomrule
	\end{tabular}
	\begin{flushleft}
		{Individual detections are highlighted in bold.}
	\end{flushleft}
	\label{tab:gals}
\end{table*}

The property {\fout} remains a useful empirical measurement for estimating the ionizing photon budget with knowledge of the UV luminosity function. In order to estimate  the quantity more closely tied to reionization models, {\fesc}, we model the underlying stellar populations of these galaxies as well as the structure of the neutral-phase inter-stellar medium (ISM). Following \citet{Steidel2018} and \citet{Pahl2021}, we use stellar-population synthesis models from BPASS \citep[BPASS v2.1,][]{Eldridge2017} together with the SMC attenuation relation  \citep{Gordon2003} at a range of $E(B-V)$ to estimate the intrinsic production rate of ionizing photons, together with geometric ISM modeling according to the ``holes" model \citep[also see][]{Reddy2016b,Reddy2021}. While the free parameters in the fit include continuum reddening in the foreground gas $E(B-V)_{\rm cov}$,  \ion{H}{I} column density log($N_{\rm HI}$/cm$^{-2}$), and covering fraction of neutral gas and dust $f_{\rm c}$, in this work we are specifically interested in {\fesc}, defined as $\fesc=1-f_{\rm c}$ for the ``holes" ISM model. Best-fit values of $\langle\fesc\rangle=0.120\pm0.009$ for {\sfrdl} and $\langle\fesc\rangle=0.122\pm0.010$ for {\sfrdu} are shown as green points in Figure \ref{fig:fesc}, presented as a function of median {\sfrd} (see right panel of Figure \ref{fig:sfrd}). Here, the {\fesc} estimates assuming an SMC attenuation relation and third-quartile IGM+CGM transparency are adopted as fiducial. The estimates of {\fesc} and {\fout} for our fiducial model are also provided in Table \ref{tab:prop} for the {\sfrdl} and {\sfrdu} composites.

We do not find a significant trend of {\fesc} with {\sfrd} within the KLCS $HST$ sample. We compare this trend to the {\fesc}-{\sfrd} relation from ``Model II" of \citet{Naidu2020}. This model used Copernicus complexio Low Resolution (\textsc{color}) dark-matter simulations \citep{Sawala2016,Hellwing2016} together with assumptions of SFR \citep{Tacchella2018} and a relation between galaxy size and halo size, and was constrained using a variety of observational data on reionization including the contaminated $\fesc=0.09\pm0.01$ measurement from \citet{Steidel2018}. This relation from \citet{Naidu2020} is $f_{esc}$ = 1.6$\Sigma_{\rm SFR}^{0.4}$. We display this relation as a black dashed line in Figure {\ref{fig:fesc}. Despite the elevated {\fesc} measurements drawn from the KLCS $HST$ sample compared to an unbiased KLCS, our {\sfrdu} measurement of {\fesc} falls significantly below the value predicted by the \citet{Naidu2020} relation for the same {\sfrd}. This 10$\sigma$ deviation may be caused by low-redshift interlopers contributing to an elevated {\fesc} from \citet{Steidel2018}, which \citet{Naidu2020} used as a constraint for the model fit, however the difference between the \citet{Steidel2018} {\fesc} and the uncontaminated measurement from \citet{Pahl2021} is only 30\%. 

We additionally measure {\fesc} using an alternative attenuation relation and average (as opposed to third-quartile) transparency of the IGM+CGM. These results, including attenuation relations from \citet{Reddy2016a}, are displayed alongside the fiducial model in Figure \ref{fig:fesc}. This second set of assumptions also does not result in a significant trend in {\fesc} across the two {\sfrd} bins.

While the measurements of the ionizing spectral region do not indicate a dependence of ionizing photon escape on {\sfrd}, the longer rest-UV wavelength spectral features in the lower panels of Figure \ref{fig:spec} appear qualitatively distinct for the two bins in {\sfrd}. We measure the equivalent widths of prominent rest-UV features such as Ly$\alpha$, low-ionization interstellar absorption lines (Si~\textsc{ii}$\lambda1260$, O~\textsc{i}$\lambda1302+$Si~\textsc{ii}$\lambda1304$, C~\textsc{ii}$\lambda1334$, and Si~\textsc{ii}$\lambda1527$), and high-ionization interstellar absorption lines (Si~\textsc{iv}$\lambda\lambda1393,1402$ and C~\textsc{iv}$\lambda1548,1550$) following the procedure of \citet{Pahl2020a} and \citet{Kornei2010}. These methods bootstrap each subsample 500 times with replacement, then perturb each individual spectrum at each pixel by a Gaussian with standard deviation set by the error spectrum. Thus, the error bars on the equivalent width measurements include both sample variance and measurement errors. We measure the Ly$\alpha$ equivalent width, {\wlya}, using the ``emission" method from \citet{Kornei2010}\footnote{Ly$\alpha$ flux was integrated between the wavelength values that the Ly$\alpha$ profile intersected the blue-side continuum value (average flux density between $1220\textrm{\AA}-1180\textrm{\AA}$) and the red-side continuum value (average flux density between $1225\textrm{\AA}-1255\textrm{\AA}$).}, and measure absorption line equivalent widths by integrating the absorption profile between $\pm5${\AA} of the line center after performing continuum normalization using the \textsc{iraf} routine \textit{continuum}. 
We present the equivalent-width measurements for both composites in Table \ref{tab:prop}. {\wlya} is elevated in the {\sfrdu} composite compared to the {\sfrdl} composite, while {\wlis} decreases and {\whis} increase. However, none of the differences is significant. When the bootstrapping is removed from the error determinations to test the contribution of sample variance, the elevation in {\wlya} and {\whis} in {\sfrdu} become significant. The lack of correlation between {\fesc} and {\sfrd}, despite the trends in {\wlya}, as well as the large sample variance in the equivalent width measurements, motivate a closer examination of how representative the KLCS $HST$ sample is and its ability to recover a significant relation between {\fesc} and {\sfrd}.

\begin{figure*} 
	\centering
	\includegraphics[width=\textwidth]{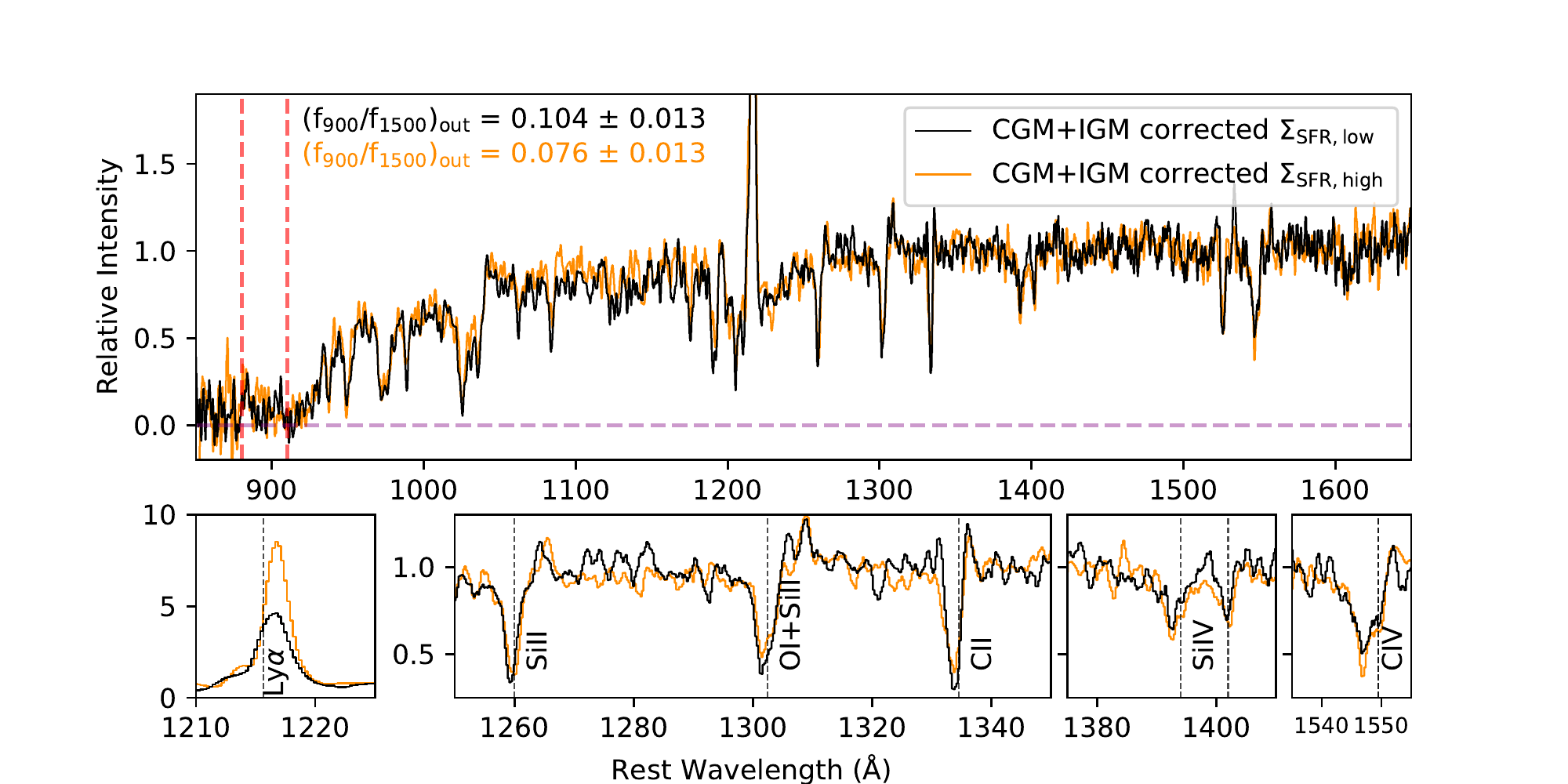}
	\caption{Composite spectra of the two KLCS $HST$ subsamples binned in {\sfrd}. The {\sfrdl} composite, containing objects with the lowest {\sfrd}, is displayed in black, while {\sfrdu}, containing objects with the highest {\sfrd}, is displayed in orange. The LyC spectral region, which defines the $f_{900}$ measurement, is within the dashed, red lines. 
		\textbf{Top:} The full composites, normalized to $f_{1500}$.
		\textbf{Bottom:} The same composites as above with wavelength and intensity limits set to highlight specific spectral features. From left to right, the panels highlight Ly$\alpha$; Si~\textsc{ii}$\lambda1260$, O~\textsc{i}$\lambda1302+$Si~\textsc{ii}$\lambda1304$ and C~\textsc{ii}$\lambda1334$; Si~\textsc{iv}$\lambda\lambda1393,1402$; and C~\textsc{iv}$\lambda1548,1550$.
	}
	\label{fig:spec}
\end{figure*}

\begin{figure} 
	\centering
	\includegraphics[width=\columnwidth]{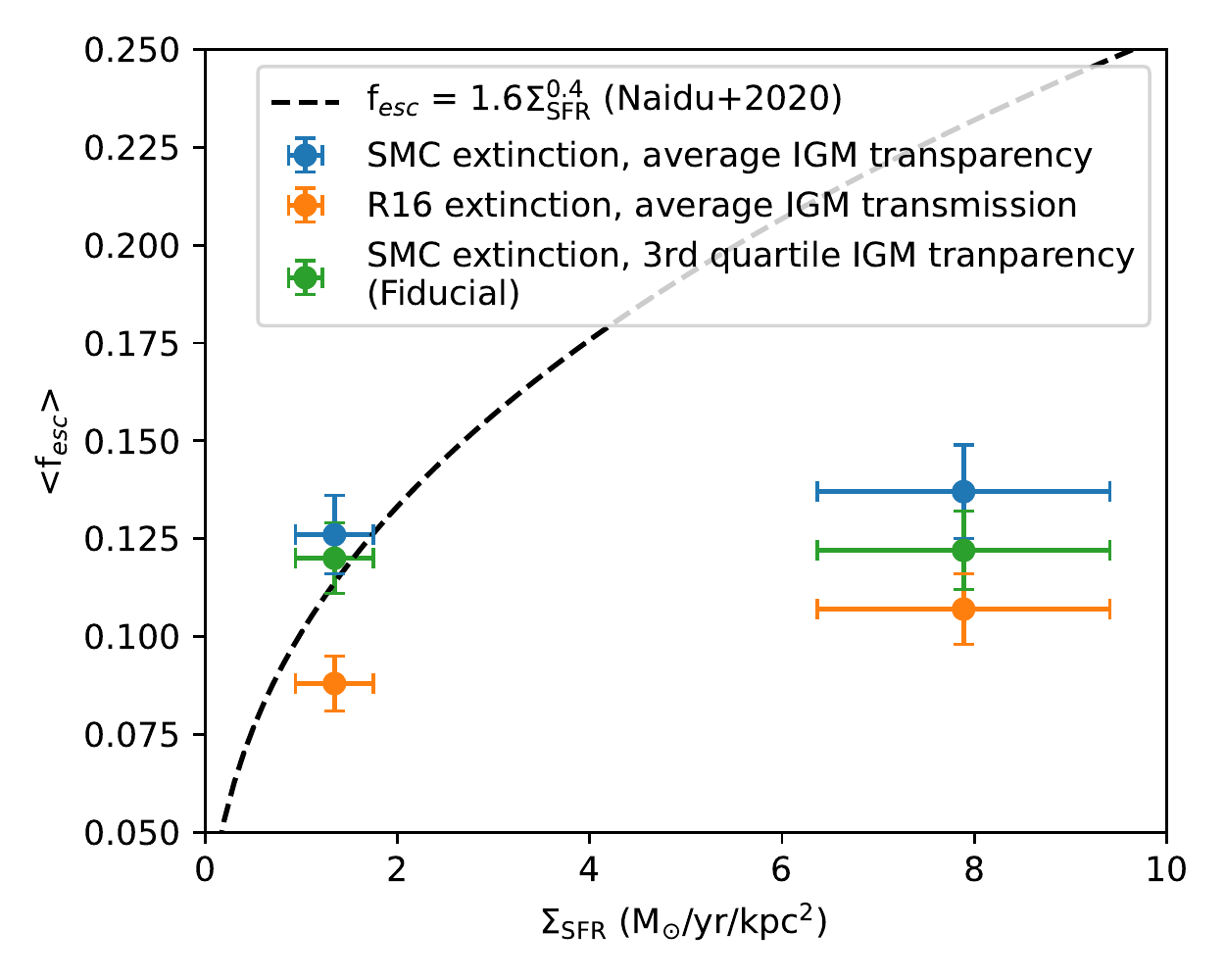}
	\caption{{\fesc} measurements for the two {\sfrd} composite spectra. The measurements assuming SMC extinction and third quartile IGM+CGM transparency, our fiducial model, are displayed in green. Additional measurements including a \citet{Reddy2016a} attenuation relation and average IGM+CGM transparency are shown as blue and orange trend lines. The {\fesc}-{\sfrd} relation from ``Model II" of \citet{Naidu2020} is shown as a dashed, black line.
	}
	\label{fig:fesc}
\end{figure}

\section{Discussion} \label{sec:disc}

In this section, we judge our ability to recover trends between galaxy properties and ionizing photon escape using the KLCS $HST$ sample, and perform simulations to determine the minimum sample size for performing these types of analyses.

{\wlya} is a well-established tracer of LyC escape in $z\sim3$ galaxies. In the full KLCS sample, both {\fout} and {\fesc} are strongly correlated with {\wlya}, measured across four equal-sized bins of {\wlya} \citep{Steidel2018,Pahl2021}. This correlation has also been found in other LyC surveys at $z\sim3-4$ \citep{Marchi2017,Marchi2018,Fletcher2019}, and is motivated physically by the similar modulation of escaping LyC and Ly$\alpha$ photons by the neutral gas covering fraction \citep{Reddy2016b,Gazagnes2020,Steidel2018}. ``Holes" in the ISM and CGM, free of \ion{H}{I} and dust, would provide clear channels for both LyC and Ly$\alpha$ photons to escape in the observer's direction. Nonetheless, Ly$\alpha$ photons may also escape off-resonance, explaining the scatter in the ratio of escaping Ly$\alpha$ to escaping LyC.
For these reasons, we use the relationship between {\wlya} and {\fout} as an indicator of the robustness of the KLCS $HST$ sample, such that, if we can recover the correlations between {\fout} and {\wlya}, the composition and size of the sample is appropriate for measuring the correlation between {\fesc} and {\sfrd} presented in Section \ref{sec:res}.

We split the KLCS $HST$ sample into two bins of {\wlya}, identical in size to those of {\sfrdl} and {\sfrdu}. {\wlyal} contains the 17 objects with the lowest {\wlya} measurements ({\wlya}$_{\rm med}=-7.0\pm3.8$), and {\wlyau} contains the 18 objects with the highest {\wlya} measurements ({\wlya}$_{\rm med}=15.8\pm6.1$). We constructed composite spectra and performed corrections for the effects of the IGM+CGM as described in Section \ref{sec:res}. The resulting {\fout} measurements are $0.11\pm0.02$ and $0.11\pm0.02$ for {\wlyal} and {\wlyau}, respectively. This lack of correlation between {\fout} and {\wlya} indicates that the KLCS $HST$ sample is not capable of recovering well-established relationships between LyC escape and key galaxy properties (i.e., {\wlya}). We repeat this experiment with two bins of increasing {\luv}, which is found to be strongly anti-correlated with {\fout} in the full KLCS \citep{Steidel2018,Pahl2021}. We measured $\fout=0.12\pm0.02$ from the composite made of the 17 galaxies with the lowest {\luv}, and $\fout=0.12\pm0.02$ from the composite made of the 18 galaxies with the highest {\luv}, supporting the assertion that the KLCS $HST$ sample is not suitable for performing correlation analysis with ionizing-photon escape and galaxy property. Finally, we bin the sample as a function of {\re}, as a correlation between LyC leakage and rest-UV size has been found in a sample of 201 galaxies from the Vimos Ultra Deep Survey at $3.5\leq z\leq4.3$ \citep{Marchi2018}. We measured $\fout=0.11\pm0.02$ and $\fout=0.11\pm0.02$ from the lower- and higher-{\re} composites, respectively, finding no significant correlation between {\re} and {\fout} in the KLCS $HST$ sample.

The $HST$ program that defines the KLCS $HST$ sample was designed to cover all 15 galaxies that were candidates for individual LyC detections in KLCS \citep{Steidel2018}. These $HST$ observations yielded 22 additional KLCS galaxies with individual LyC non-detections that fell within the footprint of the $HST$/ACS and WFC3 pointings (see Section \ref{sec:sample}). Given that these LyC non-detections are included in KLCS $HST$ only on the basis of their proximity to individual apparent LyC detections, they should be representative of the population of 106 galaxies with LyC non-detections in the full sample. Briefly, we consider the sample properties of the 22 individual non-detections in KLCS $HST$ compared to the 84 not covered by $HST$ imaging. The median {\wlya} of the KLCS $HST$ non-detections is $4.9\pm2.8$\AA, while the median {\wlya} of the non-$HST$ non-detections is $3.7\pm2.9$\AA. The $HST$ subsample does not appear different from the parent sample in {\wlya}, neither in sample median nor sample distribution. The median $\luv/\luv^*$ of the $HST$ and non-$HST$ non-detections are $0.7\pm0.1$ and $1.04\pm0.04$, respectively. The galaxies with individual LyC non-detections observed by $HST$ appear significantly fainter than the parent sample, which could affect the dynamic range of galaxy properties that the sample is probing. At the same time, the lower {\luv} of the $HST$ non detections means the constraints on {\fout} (due to the lower average $f_{1500}$) are weaker for the $HST$ sample. Additionally, the LyC individual detections have a median $\luv/\luv^*=0.88\pm0.14$, fainter than that of the objects not covered by $HST$ imaging. This difference in the individual LyC detections is expected considering the inverse correlation with {\fout} and {\luv} found in the full sample, but further contributes to KLCS $HST$ representing a biased sampling of KLCS in {\luv}. Upon stacking the $HST$ non-detections, we find the resulting {\fout} is comparable to that of the non-$HST$ non-detections, despite the inverse correlation between {\fout} and {\luv}. In addition to being systematically fainter, the KLCS $HST$ sample does not appear to follow the canonical relationship between {\fout} and {\luv}.

Despite the failure of KLCS $HST$ to recover established trends with {\fout} and galaxy properties due to its sample construction, we can still use the full KLCS sample to understand how to design future observing programs of sufficient size to recover robust trends. To inform future observing programs that select a subset of the KLCS sample to analyze, we created simulations that explore the feasibility of recovering trends between {\fout} and {\wlya} with a range of sample sizes. Here, we assume that a trend between {\fout} and {\wlya} must be recovered in order to confidently constrain trends between {\fout} other galaxy properties, considering the robustness of the {\fout}-{\wlya} connection in large-scale LyC studies \citep{Steidel2018,Gazagnes2020}. If a {\fout} and {\wlya} trend is not recovered, the sample is unlikely to be able to make confident claims about how {\fout} depends on other variables.

In the following set of simulations, we assumed that the 13 objects individually detected in LyC are always included in the analysis, similar to the design of the $HST$ program described in \citet{Pahl2021}. We began by randomly selecting a number of objects $n$ with individual LyC non-detections from KLCS. We added all 13 confirmed LyC individual detections to the sample, then split the sample into two equal bins of increasing {\wlya}, with a bin size of $(13+n)/2$.\footnote{In the cases where the total sample size was odd, the higher {\wlya} bin had one more object than the lower {\wlya} bin.} After constructing composite spectra, we performed IGM+CGM corrections using average IGM transparency for simplicity, then measured {\fout} from the IGM+CGM corrected composites. We repeated this experiment 1000 times for different random draws of $n$ galaxies with individual LyC non-detections in KLCS, always including the 13 confirmed LyC individual detections. For each iteration, we determined whether the subsequent {\fout} measurements recovered a significant increase in {\fout} across the two bins of increasing {\wlya} as reported in \citet{Steidel2018} and \citet{Pahl2021}. A $1\sigma$ increase was defined as 
\begin{equation} \label{eqn:diff}
	(\langle f_{900}/f_{1500}\rangle_{\rm out,1}-\langle f_{900}/f_{1500}\rangle_{\rm out,2})>\sqrt{(\sigma_{fout,1})^2 + (\sigma_{fout,2})^2},
\end{equation}
where the subscript ``1" denotes measurements associated with the {\wlyau} bin and the subscript ``2" denotes measurements of the {\wlyal} bin, and $\sigma_{fout}$ is the error on the corresponding {\fout} measurement.
If this condition was met, we counted the trial as a ``success." The proportion of trials that successfully recovered a 1$\sigma$ increase in {\fout} for a range of $n$ is presented as the blue, dashed curve in the upper panel of Figure \ref{fig:sim}. For $n=22$, identical to the number of non-detections included in KLCS $HST$ (and plotted as a vertical, purple, dashed line), a 1$\sigma$ increase in {\fout} across two bins of increasing {\wlya} was found in 86.6\% of random draws. In order to ensure that a 1$\sigma$ increase is recovered >95\% of the time, $n=45$ was required, for a total sample size of 58. For each iteration, we also measured the ratio of ionizing to non-ionizing flux density before IGM+CGM corrections, or {\fobs}, to explore the effects of the uncertainty of the IGM and CGM corrections on the simulations. The percentage of trials that successfully measured an increase in {\fobs} is presented as a function of $n$ in the red, dashed line in Figure \ref{fig:sim}. As expected, because of the reduced uncertainty, removing corrections for the attenuation of the IGM+CGM induced ``significant" differences in {\fobs} across the two {\wlya} bins more often. We additionally include solid curves in Figure \ref{fig:sim} that represent identical simulations with a stricter success threshold of 2$\sigma$ (i.e., a difference in {\fout} for the two bins that exceeds twice the error on the difference). For $n=22$, a 2$\sigma$ increase in {\fout} was recovered for 49.3\% of the iterations, while $n=91$ (total sample size of 104) was required for a 2$\sigma$ increase to be recovered for >95\% of iterations.

We performed a similar set of simulations without the stipulation that objects with individual LyC detections in the KLCS are preferentially observed. In this case, the subsamples for analysis were drawn randomly from the 120 galaxies in the full KLCS sample. This set of simulations is comparable to a survey that is blind to the ionizing properties of the galaxies before analysis and stacking. Here, we chose $n$ galaxies from KLCS such that $n$ is the complete sample size of the subsample. The other steps of this set of simulations were identical to those described above, including the binning as a function of {\wlya}, generation of composite spectra, correcting for the IGM+CGM, and measuring {\fout}. The percentage of trials that recovered a 1$\sigma$ increase in {\fout} across the two bins of increasing {\wlya} is presented as a function of $n$ using a dashed, blue curve in the bottom panel of Figure \ref{fig:sim}. Here, $n=35$ would be comparable to an KLCS $HST$-like program of 13 LyC detections and 22 LyC non-detections, but in this case the individual LyC detections are not oversampled. For $n=35$, a 1$\sigma$ increase in {\fout} across the two bins of increasing {\wlya} was only seen in 53.7\% of the iterations. The difference between this set of simulations and the preceding one indicates that preferentially observing LyC detections increases the ability to measure trends with {\fout} and galaxy property at fixed sample size. To recover a 1$\sigma$ difference in 95\% of trials, $n=90$ objects were required. Similar to the upper panel, we also include curves that test correlation of {\fobs} with {\wlya} and a stricter 2$\sigma$ calculation of trial success.

In summary, despite the tendency for KLCS $HST$-like samples to recover expected trends between {\fout} and {\wlya}, we do not find a significant correlation between {\fout} and {\wlya} in KLCS $HST$. This result is somewhat unexpected, considering this lack of correlation is only recovered in $\sim13\%$ of trials at similar sample size. The fact that we cannot recover the connection between {\fout} and {\wlya} in the KLCS $HST$ sample calls into question the robustness of the lack of correlation found between {\fesc} and {\sfrd} in Section \ref{sec:res}. Our simulations show that we require a larger sample size to ensure that a significant correlation is recovered between {\fout} and {\wlya}, which we have assumed is a requirement to discern a significant trend between {\fesc} and {\sfrd}. Accordingly, a larger sample size is required to make robust statements about the connection between {\fesc} and {\sfrd}. The simulations presented in this section can be used to dictate future $HST$ LyC observing programs. However, these results assume a parent sample with spectra of similar quality to KLCS, and having the same intrinsic correlation. With better quality spectra or a sample that covers a larger dynamic range of galaxy property (e.g., {\sfrd}, {\luv}), the necessary sample size may decrease. Nonetheless, these results provide insight into how measurement uncertainty, IGM sightline variability, and sample size can affect eventual measurements of the connection between  {\fesc} and galaxy properties.

\begin{figure} 
	\centering
	\includegraphics[width=\columnwidth]{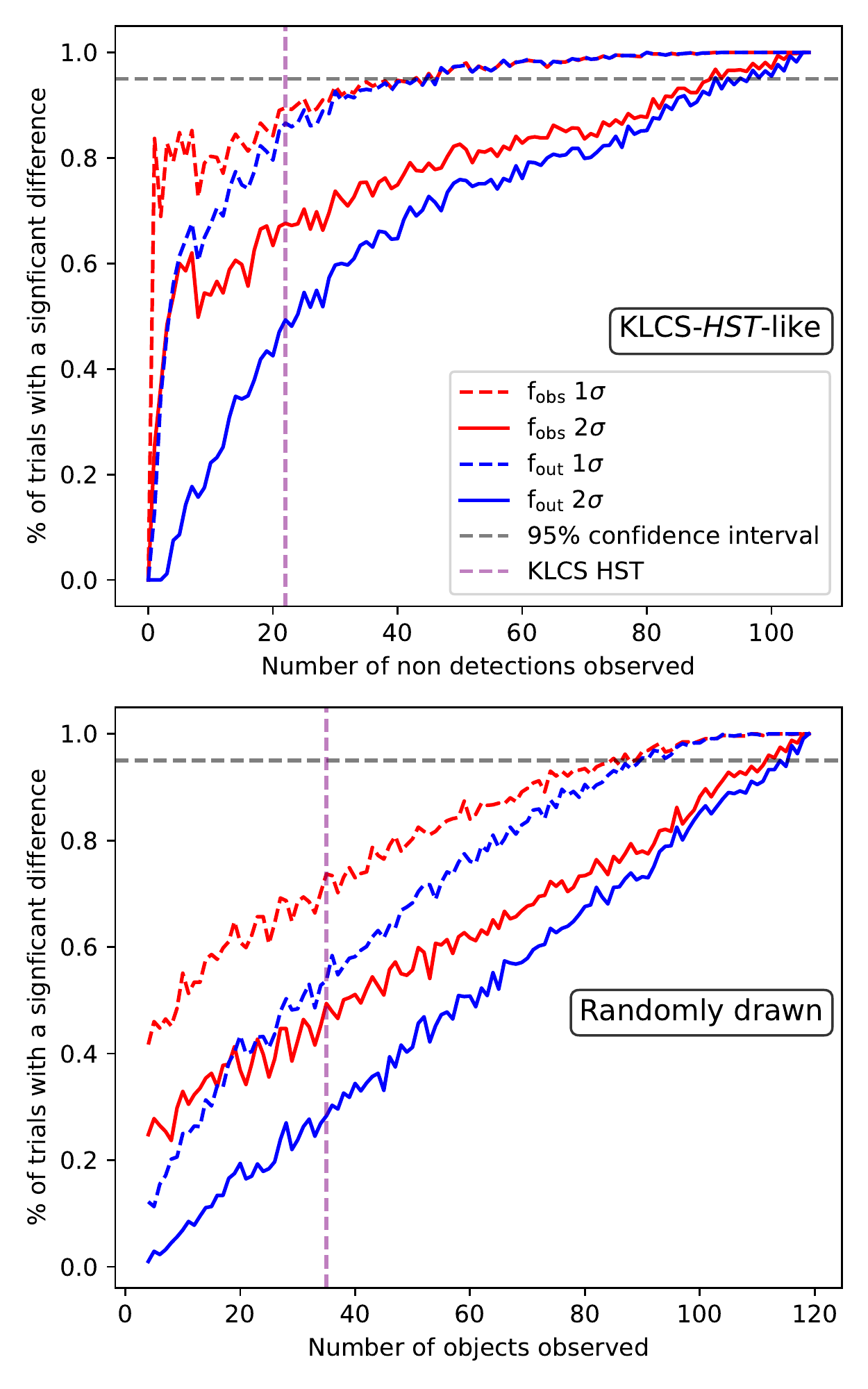}
	\caption{Simulations to determine if a subsample of a given size is sufficient to recover an increase of \ensuremath{\langle f_{900}/f_{1500}\rangle} across two equal bins of {\wlya}. A successful trial was defined as a case in which the \ensuremath{\langle f_{900}/f_{1500}\rangle} measurement of the upper {\wlya} composite was greater than the corresponding measurement from the lower {\wlya} composite at a 1$\sigma$ confidence level for the dashed lines, and 2$\sigma$ for the solid lines. The blue curves represent the likelihood of significant correlations recovered for {\fout}, which includes corrections for attenuation from the IGM+CGM attenuation, while the red curves use simply {\fobs}. For each sample size, 1000 random draws were performed to determine the percent chance of success.
		\textbf{Top:} KLCS-$HST$-like sample construction, which includes 13 individually-detected LyC objects for every trial and a number of \textit{n} non-detections, for a total sample size of $13+n$. The number of non-detections included in KLCS $HST$ is 22, represented by a purple, vertical, dashed line.
		\textbf{Bottom:} Sample construction includes $n$ randomly drawn galaxies from KLCS. Here, KLCS $HST$ is represented by a purple, vertical, dashed line at 35, ignoring the oversampling of individually-detected objects in KLCS $HST$.
	}
	\label{fig:sim}
\end{figure}

\section{Summary} \label{sec:summary}

The timeline of reionization remains uncertain, as the main sources that drive the process are still debated. Estimating {\fesc} in samples of lower-redshift analogues and studying how their properties connect with {\fesc} is a fruitful endeavor to distinguish among various models of reionization. In order to explore the potential connection between {\fesc} and {\sfrd}, which is well motivated by the physics of star-formation feedback \citep[e.g.,][]{Sharma2016,Ma2016,Kakiichi2021}, we examined a sample of 35 galaxies drawn from the KLCS survey which have been shown to be free of line-of-sight contamination based on $HST$ imaging. We measured {\re} using Sérsic profile fits to high-resolution, {\hstv} images, and SFR from SED fits to ground- and space-based photometry. We split the sample into two bins of {\sfrd}, and constructed composite rest-UV spectra to examine whether their ionizing and non-ionizing rest-UV spectral features depend on the sample median {\sfrd}. The main results of this work are as follows:

\begin{enumerate}
	\item We measure no significant difference in {\fesc} for spectral composites constructed from the KLCS $HST$ sample split into two bins of {\sfrd}. This null result is confirmed for a range of assumptions regarding IGM transparency and dust attenuation. Redward of LyC, we measure an elevated {\wlya} and {\whis} in the composite characterized by the larger {\sfrd}, although the differences were not significant in the face of sample variance.
	\item We measure no significant difference in {\fesc} across two composites binned as a function of {\wlya}, despite {\fesc}-{\wlya} relationships found in the full KLCS, and other LyC surveys \citep[e.g.,][]{Fletcher2019,Gazagnes2020}. Repeating the experiment using two bins of {\luv}, we also find no significant difference in {\fesc} between composite spectra.
	These results indicate that the KLCS $HST$ sample is unable to robustly constrain correlations between {\fesc} and other galaxy properties, including {\sfrd}.
	\item We find that a sample of >58 galaxies (13 galaxies individually detected in LyC, 45 individually undetected) would be required to recover a significant positive change in {\fout} among in two subsamples binned by {\wlya} $>95\%$ of the time. This recommendation is based on stacking simulations using the full KLCS, randomly drawing 45 individual LyC non-detections while always including the 13 LyC detections, similar to the construction of the KLCS $HST$ sample. With a KLCS-$HST$-like sample of 13 LyC detections and 22 randomly-drawn non-detections, a significant correlation is recovered in only 86.6\% of realizations. Here, the actual KLCS $HST$ sample corresponds to one of the $\sim13\%$ of trials in which a correlation would not be recovered.
	\item We repeat the stacking simulations without the assumption that the 13 individual LyC detections are preferentially observed, and find that a sample of >90 objects is required to recover a $1\sigma$ increase in {\fout} across two bins of increasing {\wlya} in >95\% of trials. These simulations are performed by drawing 90 objects randomly from KLCS before binning and stacking, and did not include all detections in every mock sample. These simulations can be used to inform future LyC surveys aimed at studying dependencies of {\fesc} on other galaxy properties, although the recommended numbers depend on spectral SNR and dynamic range of the sample.
\end{enumerate}

We have demonstrated that the KLCS $HST$ sample is unfortunately not appropriate for recovering possible trends between {\fesc} and {\sfrd}, due to a combination of small sample size and non-representativeness in terms of the distribution of quantities such as {\luv} at fixed {\wlya}.
We have also outlined what must be done to further explore the {\fesc}-{\sfrd} connection: more objects are needed with $HST$-resolution imaging appropriate for accurate size estimates and contamination rejection, combined with deep, rest-UV spectra to estimate {\fesc} in stacked composites with sufficient objects in each bin. We will better constrain these important trends using KLCS after more individual LyC non-detections are followed up by $HST$.
\linebreak
\linebreak
We acknowledge support from NSF AAG grants 0606912, 0908805, 1313472 2009313, 2009085, and 2009278.
Support for program HST-GO-15287.001 was provided by NASA through a grant from the Space Telescope Science Institute, which is operated by the Associations of Universities for Research in Astronomy, Incorporated, under NASA contract NAS5-26555. 
CS was supported in part by the Caltech/JPL President's and Director's program.
Based on observations obtained with MegaPrime/MegaCam, a joint project of CFHT and CEA/IRFU,
at the Canada-France-Hawaii Telescope (CFHT) which is operated by the National Research Council (NRC) of Canada,
the Institut National des Science de l'Univers of the Centre National de la Recherche Scientifique (CNRS) of France,
and the University of Hawaii. This work is based in part on data products produced at Terapix available at the
Canadian Astronomy Data Centre as part of the Canada-France-Hawaii Telescope Legacy Survey, a collaborative project of NRC and CNRS.
We wish to extend special thanks to those of Hawaiian ancestry on
whose sacred mountain we are privileged to be guests. Without their generous hospitality, most
of the observations presented herein would not have been possible.

\section*{Data Availability Statement}
The {\it HST} data presented in this article are publicly available
from the Mikulski Archive for Space Telescopes. The ground-based
data presented here will be shared on reasonable request to the
corresponding author.


\end{document}